\documentclass[showpacs,showkeys,preprintnumbers,pre,amsmath,amssymb,superscriptaddress,twocolumn]{revtex4}
\usepackage[english]{babel}
\usepackage[utf8]{inputenc}
\usepackage[colorinlistoftodos, color=green!40, prependcaption]{todonotes}

\usepackage[colorlinks=true,linkcolor=blue,urlcolor=blue,citecolor=blue]{hyperref}

\begin{document}
\title{A Mesoscale Perspective on the Tolman Length}

\author{Matteo Lulli}
    \email[Correspondence email address: ]{lulli@sustech.edu.cn}
    \affiliation{Department of Mechanics and Aerospace Engineering, Southern University of Science and Technology, Shenzhen, Guangdong 518055, China}
    
\author{Luca Biferale}
    \email{biferale@roma2.infn.it}
    \affiliation{Department of Physics \& INFN, University of Rome ``Tor Vergata'', Via della Ricerca Scientifica 1, 00133, Rome, Italy.}

\author{Giacomo Falcucci}
    \email{giacomo.falcucci@uniroma2.it}
    \affiliation{Department of Enterprise Engineering ``Mario Lucertini'', University of Rome ``Tor Vergata", Via del Politecnico 1, 00133 Rome, Italy; John A. Paulson School of Engineering and Applied Physics, {\it Harvard University},  33 Oxford Street, 02138 Cambridge, Massachusetts, USA.}

\author{Mauro Sbragaglia}
    \email{sbragaglia@roma2.infn.it}
    \affiliation{Department of Physics \& INFN, University of Rome ``Tor Vergata'', Via della Ricerca Scientifica 1, 00133, Rome, Italy.}

\author{Xiaowen Shan}
    \email{shanxw@sustech.edu.cn}
    \affiliation{Department of Mechanics and Aerospace Engineering, Southern University of Science and Technology, Shenzhen, Guangdong 518055, China}

\date{\today} 

\begin{abstract}
    We demonstrate that the multi-phase Shan-Chen lattice Boltzmann method (LBM) yields a curvature dependent surface tension $\sigma$ as computed from three-dimensional hydrostatic droplets/bubbles simulations. Such curvature dependence is routinely characterized, at first order, by the so-called {\it Tolman length} $\delta$. LBM allows to precisely compute $\sigma$ at the surface of tension $R_s$ and determine the Tolman length from the coefficient of the first order correction. The corresponding values of $\delta$ display universality for different equations of state, following a power-law scaling near the critical temperature. The Tolman length has been studied so far mainly via computationally demanding molecular dynamics (MD) simulations or by means of density functional theory (DFT) approaches playing a pivotal role in extending Classical Nucleation Theory. The present results open a new hydrodynamic-compliant mesoscale arena, in which the fundamental role of the Tolman length, alongside real-world applications to cavitation phenomena, can be effectively tackled. All the results can be independently reproduced through the ``idea.deploy" framework.
\end{abstract}
\maketitle

\section{Introduction}\label{sec:intro}
\begin{figure}[t!]
    \includegraphics[scale = 0.485]{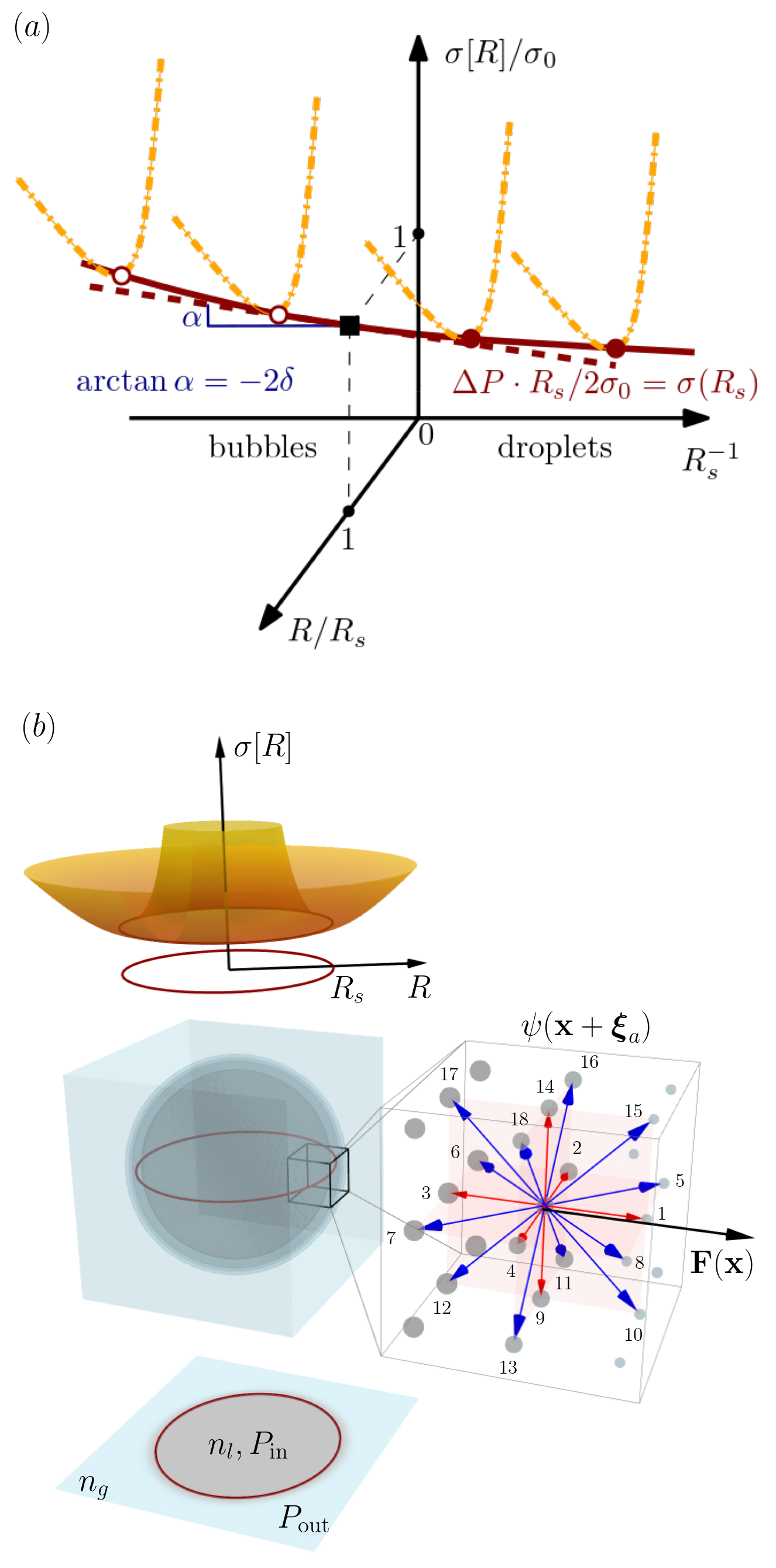}
    \caption{Panel $(a)$: Sketch of the generalized surface tension $\sigma[R]$ normalized to the flat-interface value $\sigma_0$, as a function of the droplet curvature $R_s^{-1}$ and of the normalized arbitrary dividing surface $R/R_s$. The Tolman length $\delta$ is given by the slope of the locus of the minima of $\sigma[R]$ in the flat interface limit $R_s^{-1}\to 0$. Panel $(b)$. Top: data for the generalized surface tension $\sigma[R]$ with the minima determining the surface of tension $R_s$. Middle part: density field $n$ of a droplet with an enhancement of the underlying lattice structure with the discrete velocities reported in red and blue for $|\boldsymbol{\xi}_a|^2 = 1,2$ respectively. The points shading and size corresponds to the magnitude of $\psi$ at the interface with $\mathbf{F}$ the local force as in Eq.~\eqref{eq:SCForce}. In the bottom projection of the density field $n$ we indicate the positions for the inner $P_{\text{in}}$ and outer $P_{\text{out}}$ bulk pressures together with the liquid $n_l$ and gas $n_g$ densities for a droplet.}
    \label{fig:sketch}
\end{figure}

Defining the position of the interface in a multi-phase mixture is not a straightforward task. Given a bubble/droplet, the average density profile changes smoothly and not as a step-wise function, so that the exact position of a surface separating the two phases is an elusive concept. On the other hand, for closed interfaces, the curvature appears explicitly in the free energy as conjugated to a curvature coefficient, i.e., the curvature plays the role of a control parameter~\cite{GibbsCollected1948, Buff1951}. In this context, the introduction of an \emph{arbitrary dividing surface}, ideally separating the gas and the liquid phases, is found to be necessary~\cite{GibbsCollected1948, Buff1951, RowlinsonWidom82}. The arbitrariness of the location $R$ of such an interface does not impact on the value of the free energy, i.e. the free energy is stationary with respect to variations of $R$. This, in turn, reflects on the definition of a \emph{generalized} surface tension $\sigma[R]$, which assumes the shape of a convex function reaching a minimum at $R_s$ (see Fig.~\ref{fig:sketch}), identifying the \emph{surface of tension}. At the latter position the Laplace law applies in the usual form~\cite{GibbsCollected1948, Buff1951, RowlinsonWidom82, Blokhuis1992}. It is possible to show~\cite{RowlinsonWidom82} that the stationarity of the free energy at the surface of tension $R_s$ yields
\begin{equation}\label{eq:f-statio-Rs}
    \Delta P = \frac{2\sigma[R_s]}{R_s} + \frac{\mbox{d}\sigma[R]}{\mbox{d}R}\bigg|_{R = R_s} = \frac{2\sigma(R_s)}{R_s}.
\end{equation}
By considering any other value of $R\neq R_s$ in~\eqref{eq:f-statio-Rs}, such as the equimolar radius $R_e$ commonly used in LBM simulations, one obtains the so-called \emph{generalized} Laplace law which explicitly depends on the derivative of $\sigma[R]$.
The locus of the minima of $\sigma[R]$ identifies a physical, i.e. non-arbitrary, dependence of the surface tension $\sigma(R)$ on the droplet/bubble size at $R_s$, $\sigma_s = \sigma[R_s] = \sigma(R_s)$. Such a dependence was first examined in the seminal paper by Tolman~\cite{Tolman1949} (see~\cite{Malijevsky12, Ghoufi_2016} for reviews), and can be expressed as a power-law expansion in the curvature, i.e. the inverse radius, which at second order reads~\cite{Blokhuis1992, Blokhuis1992Rigidity, Aasen2018, Rehner2019}
\begin{equation}\label{eq:sigma_c}
\sigma\left(R_s\right)\simeq\sigma_{0}\left(1-\frac{2\delta}{R_s}+\frac{2\bar{k}+k}{R_s^{2}}\right).
\end{equation}
The flat interface value $\sigma_0$ appears at the leading order, the first order coefficient $\delta$ defines the Tolman length [cf. Fig.~\ref{fig:sketch}$(a)$] and $\bar{k}$ and $k$ are called curvature and Gaussian-rigidity coefficients, respectively. The present work mainly focuses on the analysis of $\delta$ since, as shown in the results, the higher order coefficients $k$ and $\bar{k}$ are small enough to make higher order terms negligible in the present setting. 

In Tolman's seminal work~\cite{Tolman1949} $\delta$ was defined on thermodynamic grounds, starting from Gibbs theory of capillarity~\cite{GibbsCollected1948}. Such an approach, further developed in~\cite{Henderson1984, Bartell2001, Blokhuis2006}, served as the foundation for studying the behavior of $\delta$ near the critical temperature~\cite{Rowlinson1984, Blokhuis1992, Anisimov2007}. Several works based on the density functional theory (DFT)~\cite{Blokhuis2013, Wilhelmsen2015, Aasen2018, Rehner2019} have led to expressions for the coefficients $\delta$, $k$ and $\bar{k}$, for realistic multi-phase and multi-component systems. From the numerical perspective, simulations have mostly focused on molecular dynamics (MD)~\cite{Nijmeijer1992, VanGiessen2009, Menzl2016}; Monte Carlo techniques have been adopted as well, as in the three-dimensional three-body Ising model~\cite{Troster2011} and for particles interacting via Lennard-Jones (LJ) potentials~\cite{Rao1979, TenWolde1998, Moody_2001}.
The Tolman length was recently investigated in experimental settings, linked to hydrophobic interactions relevant for protein folding~\cite{Chen_2007}; $\delta$ was measured in nucleation experiments~\cite{Bruot2016}, and its role was analyzed both in confined geometries~\cite{Kim2018} and in colloidal liquids~\cite{Nguyen2018}.
Corrections to the zero-curvature value $\sigma_0$ have important physical consequences, most notably regarding Classical Nucleation Theory (CNT). The latter states that, using the so-called~\emph{capillary approximation}, the nucleation rate depends exponentially on $\sigma_0$~\cite{Debenedetti1996}. Hence, such rates are extremely sensitive to curvature corrections. The latter have been successfully used to extend CNT~\cite{Talanquer1995, Tanaka_2015} and for the analysis of experimental data~\cite{Bruot2016, Nguyen2018}, eventually allowing to solve previous CNT controversial results~\cite{Aasen2020}.

In this work, we study the Tolman length using a three-dimensional multi-phase~\cite{ShanChen93, ShanChen94} lattice Boltzmann method (LBM)~\cite{kruger2017lattice, succi2018lattice} in the hydrostatic limit. We estimate $\delta$ by directly computing $\sigma[R]$ (see Fig.~\ref{fig:sketch}$(b)$) from a lattice formulation of the pressure tensor~\cite{Shan08} following a procedure reported in~\cite{RowlinsonWidom82} which we detail below. Considering the past literature it appears that, so far, the different approaches for modelling and study the Tolman length have been mainly concerned either with the microscopic scales, i.e. MD simulations, or with continuum DFT descriptions. Indeed, a mesoscale perspective has been considered in the MC simulations of the Ising model~\cite{Block2010, Troster2011, Binder2012, Binder2016a}, which however do not naturally extend to non-equilibrium settings. Here we present a first step for a mesoscale modelling of the Tolman length which embeds momentum conservation, i.e. hydrodynamics, thus allowing to consider non-equilibrium effects, which are paramount in non-homogeneous cavitation and nucleation, and to fill the mescoscopic gap seprating MD simulations and DFT theories.

The paper is organized as follows: in Section~\ref{sec:lbm} we describe the fundamentals of the LBM formulation adopted in this work, highlighting the existence of a lattice pressure tensor for the Shan-Chen model solving the mechanic equilibrium condition for a flat interface to machine precision, i.e. constant value of the normal component of the pressure tensor; in Section~\ref{sec:method} we detail the method used to evaluate the position of the surface of tension $R_s$ differing from the equimolar surface $R_e$~\cite{GibbsCollected1948, RowlinsonWidom82} typically used in the context LBM simulations; in Section~\ref{sec:res} we report the results for the estimation of the Tolman length $\delta$ and its temperature dependence and in Section~\ref{sec:conslusions} we draw some conclusions.

\section{Lattice Boltzmann Method}\label{sec:lbm}
The lattice Boltzmann method (LBM) allows to simulate the Navier-Stokes dynamics of a multi-phase mixture by means of a forced Boltzmann transport equation acting on a discretized phase-space~\cite{kruger2017lattice, succi2018lattice}: the single-particle distribution function $f(\mathbf{x}, \boldsymbol{\xi}, t)$ takes values on the the nodes $\{\textbf{x}\}$ of a three-dimensional lattice at discrete times $t$. The key advantage of LBM lays in a remarkably fast convergence to the hydrodynamic limit by employing only a few velocity vectors $\{\boldsymbol{\xi}_i\}$ connecting each lattice point to a set neighboring nodes, with $i=0,\ldots,18$. Hence, one defines the \emph{populations} as the single-particle distribution function evaluated for a given $\boldsymbol{\xi}_i$, i.e. $f_i(\textbf{x}, t) = f(\textbf{x},\boldsymbol{\xi}_i,t)$. The first two moments of the discretized distribution define the density $n = \sum_i f_i$ and the momentum density $n\textbf{u} = \sum_i \boldsymbol{\xi}_i f_i$, respectively. The lattice transport equation reads
\begin{equation}\label{eq:lbm}
    f_{i}\left(\mathbf{x}+\boldsymbol{\xi}_{i},t+1\right)-f_{i}\left(\mathbf{x},t\right)=\Omega_{i}\left(\mathbf{x},t\right) + F_i(\textbf{x}, t),
\end{equation}
where $F_i$ is the forcing term~\cite{Guo2002} and $\Omega_i$ is the local collision operator conserving mass and momentum, i.e. $\sum_i \Omega_i = \sum_i \boldsymbol{\xi}_i \Omega_i= 0$, and the locality of $\Omega_i$ renders the approach particularly amenable to parallel implementations~\cite{kruger2017lattice, succi2018lattice}, such as the architecture-independent GPU/CPU implementation used for the results reported in this paper which can be found on the GitHub repository \href{https://github.com/lullimat/idea.deploy}{https://github.com/lullimat/idea.deploy}~\cite{sympy, scipy, numpy0, numpy1, scikit-learn, matplotlib, ipython, pycuda_opencl}. The left-hand side of~\eqref{eq:lbm} represents the populations streaming while on the right-hand side is composed by the Bhatnagar-Gross-Krook (BGK)~\cite{Bhatnagar_1954} collision operator
\begin{equation}\label{eq:sm:omega_i}
    \Omega_{i}\left(\mathbf{x},t\right)=-\frac{1}{\tau}\left[f_{i}\left(\mathbf{x},t\right)-f_{i}^{\left(\text{eq}\right)}\left(\mathbf{x},t\right)\right]
\end{equation}
and by the Guo~\cite{Guo2002} forcing term
\begin{equation}
\begin{split}F_{i}\left(\mathbf{x},t\right) & =\left(1-\frac{1}{2\tau}\right)w_{i}\\
\times & \left[\frac{1}{c_{s}^{2}}\xi_{i}^{\alpha}+\frac{1}{c_{s}^{4}}\left(\xi_{i}^{\alpha}\xi_{i}^{\beta}-c_{s}^{2}\delta^{\alpha\beta}\right)u_{\beta}^{\left(\text{eq}\right)}\right]F_{\alpha},
\end{split}
\end{equation}
where repeated Greek indices imply summation. We use this term to implement in the LBM the force $F_{\alpha}$ responsible for the phase separation. The equilibrium populations $f_{i}^{\left(\text{eq}\right)}$ are obtained as a second-order approximation of the Maxwell distribution
\begin{equation}
  f_{i}^{\left(\text{eq}\right)}=w_{i}n\left[1+\frac{\xi_{i}^{\alpha}u_{\alpha}^{\left(\text{eq}\right)}}{c_{s}^{2}}+\frac{(\xi_{i}^{\alpha}u_{\alpha}^{\left(\text{eq}\right)})^{2}}{c_{s}^{4}}-\frac{u_{\alpha}^{\left(\text{eq}\right)}u_{\alpha}^{\left(\text{eq}\right)}}{2c_{s}^{2}}\right],
\end{equation}
and the equilibrium fluid velocity is computed according to Guo prescription~\cite{Guo2002}
\begin{equation}
u_{\left(\text{eq}\right)}^{\alpha}\left(\mathbf{x},t\right)=\frac{1}{n\left(\mathbf{x},t\right)}\sum_{i=0}^{18}\xi_{i}^{\alpha}f_{i}\left(\mathbf{x},t\right)+\frac{1}{2n\left(\mathbf{x},t\right)}F^{\alpha}\left(\mathbf{x},t\right).
\end{equation}
Since its inception, LBM has witnessed the development of different approaches for multi-phase flows~\cite{kruger2017lattice, succi2018lattice} laying at the foundation of the most modern and successful application of LBM. In this paper we delve deeper in one specific approach, namely the Shan-Chen (SC) model~\cite{ShanChen93, ShanChen94}, and show that it correctly captures a curvature dependent surface tension. The main feature of the SC model, allowing for the existence of stable gradients of the density $n(\textbf{x}, t)$, is a force computed on the lattice nodes
\begin{equation}\label{eq:SCForce}
    F^\mu(\mathbf{x}) = -Gc_s^2\,\psi(\mathbf{x})\sum_{a = 1}^{18} W(|\boldsymbol{\xi}_a|^2)\, \psi(\mathbf{x} + \boldsymbol{\xi}_a)\, \xi^\mu_a,
\end{equation}
where $\psi(\textbf{x}, t) = \psi(n(\textbf{x}, t))$ is the so-called pseudopotential, a local function of the density $n$, implicitly depending on space and time, $c_s = 1/\sqrt{3}$ is the speed of sound, $G$ is the (self) coupling constant which is related to the temperature, $\boldsymbol{\xi}$ are the discrete forcing directions such that their squared lengths are $|\boldsymbol{\xi}_a|^2 = 1,2$, and $W(1) = 1/6$ and $W(2) = 1/12$ are the weights ensuring 4-th order lattice force isotropy~\cite{Shan06, Sbragaglia07}.
The set of the forcing vectors ${\boldsymbol{\xi}_a}$ coincide with that of the lattice velocities ${\boldsymbol{\xi}_i}$ after excluding the ``rest" direction $\xi_0 = (0, 0, 0)$. The SC force is related to a \emph{lattice} pressure tensor~\cite{Shan08, Belardinelli15, Lulli_2021} that reads
\begin{equation}\label{eq:sm:pt}
    P^{\mu\nu}(\mathbf{x})=nc_{s}^{2}\delta^{\mu\nu}+\frac{Gc_{s}^{2}}{2}\psi(\mathbf{x})\sum_{a=1}^{18}W\left(|\boldsymbol{\xi}_{a}|^{2}\right)\psi(\mathbf{x}+\boldsymbol{\xi}_{a})\xi_{a}^{\mu}\xi_{a}^{\nu}.
\end{equation}
We remark that the tensor in the Eq.~\eqref{eq:sm:pt}  is such that the flat-interface mechanical equilibrium condition, i.e. constant normal component $P_{\text{N}}(x) = p_0$ throughout the interface, is obeyed \emph{on the lattice} with a value of $p_0$ that is constant to machine precision. By performing the Taylor expansion of Eq.~\eqref{eq:sm:pt} one obtains, at the leading order, the bulk pressure 
\begin{equation}\label{eq:bulkp}
    P(n) = n c_s^2 + \frac{Gc_s^2e_2}{2}\psi^2(n)
\end{equation} 
where $e_2 = 1$ for the values of the weights used in this work. Eq.~\eqref{eq:bulkp} allows for phase coexistence when the coupling is below the critical value $G < G_c$, which is determined by the vanishing of the first and second derivatives in $n$, i.e. $\mbox{d}P/\mbox{d}n=0$ and $\mbox{d}^2P/\mbox{d}n^2=0$. 
The SC model has been steadily developed during the past thirty years allowing to perform the most diverse simulations: from heterogeneous cavitation~\cite{Falcucci13a} to emulsions rheology~\cite{Lulli_2018}, all while handling complex boundary and load conditions, allowing for a direct comparison with microfluidics experiments~\cite{Derzsi_2018}.
The ability to model the Tolman length in LBM opens new fundamental research avenues for the study of nucleation and cavitation phenomena in the mesoscale regime, offering at the same time great computational efficiency and a direct bridge to experiments.

\section{Method}\label{sec:method}
As outlined in Section~\ref{sec:intro}, the free energy needs to be independent on the choice of the position for the arbitrary dividing spherical surface $R$. Such a stationarity condition yields the generalized Laplace law~\cite{GibbsCollected1948, Buff1951, RowlinsonWidom82, Rowlinson1984}
\begin{equation}\label{eq:LaplaceGen}
    \Delta P=\frac{2\sigma\left[R\right]}{R}+\left[\frac{\text{d}\sigma}{\text{d}R}\right]
\end{equation}
with $\sigma[R]$ the generalized surface tension and its notional derivative $[\text{d}\sigma/\text{d}R] = \sigma'[R]$ and $\Delta P = P_{\text{in}} - P_{\text{out}}$, with $P_{\text{in}}$ and $P_{\text{out}}$ the values of the bulk pressure in the center of the bubble/droplet and far away from the interface, respectively (see Fig.~\ref{fig:sketch}). At the minimum of $\sigma[R]$, Eq.~\eqref{eq:LaplaceGen} reduces to the usual Laplace law, and the condition $\sigma'[R]|_{R=R_s} = 0$ defines the position of the surface of tension $R_s$. Hence, by direct comparison to Eqs.~\eqref{eq:sigma_c} and~\eqref{eq:LaplaceGen}, it follows that at second order in $R_s^{-1}$ the latter reads
\begin{equation}\label{eq:Laplace2nd}
    \Delta P = \frac{2\sigma_s(R_s)}{R_s}\simeq \frac{2\sigma_0}{R_s}\left(1 - \frac{2\delta}{R_s}\right).
\end{equation}
In order to estimate the Tolman length we simulate droplets and bubbles at different temperature (i.e. coupling $G$) and compute the deviations from the Laplace law using the surface of tension radius $R_s$ to estimate the bubble/droplet size. We now discuss how to estimate $R_s$ from the simulations generalizing the arguments presented in~\cite{RowlinsonWidom82} to an arbitrary spatial dimension $d$. Let us start from the mechanic equilibrium condition $\partial_\mu P^{\mu\nu} = 0$ and consider the following pressure tensor decomposition
\begin{equation}\label{eq:pt_projection}
    P^{\mu\nu} = P_{\text{N}}\delta^{\mu\nu} - (P_{\text{N}} - P_{\text{T}})q^{\mu\nu},
\end{equation}
where $P_{\text{N}}$ and $P_{\text{T}}$ are the (locally) normal and tangential components to the bubble/droplet interface, respectively. The projector along the tangential direction is defined as $q^{\mu\nu}=\delta^{\mu\nu} - n^\mu n^\nu$ where $n^\mu$ is the normal vector to the interface which is given by the direction of the largest gradient. It follows that the normal vector for a droplet interface has the opposite orientation with respect to the one of a bubble, so that the latter yields a negative curvature. Hence, the mechanic equilibrium condition reads
\begin{equation}
    \partial_{\mu}P^{\mu\nu}= n^{\nu}n^{\mu}\partial_{\mu}P_{\text{N}}+n^{\nu}\partial_{\mu}n^{\mu}\left(P_{\text{N}}-P_{\text{T}}\right) = 0,
\end{equation}
which can be re-expressed in polar coordinates as
\begin{equation*}
    n^{\nu}\frac{\text{d}}{\text{d}r}P_{\text{N}}\left(r\right)+\frac{\left(d-1\right)n^{\nu}}{r}\left[P_{\text{N}}\left(r\right)-P_{\text{T}}\left(r\right)\right] =0,
\end{equation*}
where we considered that $n^{\nu}\partial_{\mu}n^{\mu}={\left(d-1\right)n^{\nu}}/{r}$, where $d$ is the number of spatial dimensions and $r$ is the value of the radial coordinate. Finally, without loss of generality, we can select the normal/radial direction to be parallel to the $x$-axis, i.e. $n^\mu = e_x^\mu$ yielding
\begin{equation}\label{eq:mech_eq_radial}
    \frac{\text{d}}{\text{d}r}P_{\text{N}}\left(r\right)+\frac{d-1}{r}\left[P_{\text{N}}\left(r\right)-P_{\text{T}}\left(r\right)\right] =0.
\end{equation}
Now, it is possible to obtain a sequence of identities that are satisfied by the mechanic equilibrium condition. As a first step one can multiply Eq.~\eqref{eq:mech_eq_radial} by $r^n$ so that, after reshuffling derivatives, one obtains
\begin{equation}
    \frac{\text{d}}{\text{d}r}[r^{n}P_{\text{N}}(r)]=r^{n-1}[(n-(d-1))P_{\text{N}}(r)+(d-1)P_{\text{T}}(r)].
\end{equation}
Next, it is possible to take the integral of both sides between $R_{\text{in}}$ and $R_{\text{out}}$, i.e. from the position of the inner bulk phase to the position of the outer bulk phase, thus obtaining
\begin{equation}\label{eq:mech_eq_integral}
    \begin{split}
        & R_{\text{out}}^{n}P_{\text{out}}-R_{\text{in}}^{n}P_{\text{in}} \\
        =&\int_{R_{\text{in}}}^{R_{\text{out}}}\text{d}r\;r^{n-1}[(n-(d-1))P_{\text{N}}(r)+(d-1)P_{\text{T}}(r)],\\
    \end{split}
\end{equation}
where we have identified the value of the normal component in the bulk with the value of the scalar pressure, $P_{\text{N}}^{\text{in,out}} = P_{\text{in,out}}$. Now, let us define the pressure-jump function $P_{\text{J}}(r;R)=P_{\text{in}}-(P_{\text{in}}-P_{\text{out}})\theta(r-R)$, where $\theta(r - R)$ is the Heaviside function. The integral of $nr^{n-1}P_{\text{J}}\left(r;R\right)$ between $R_{\text{in}}$ and $R_{\text{out}}$ reads
\begin{equation}\label{eq:p_jump_integral}
    \begin{split}
        & n\int_{R_{\text{in}}}^{R_{\text{out}}}\text{d}r\;r^{n-1}P_{\text{J}}(r;R)\\
        &=P_{\text{out}}R_{\text{out}}^{n}-P_{\text{in}}R_{\text{in}}^{n}+R^{n}(P_{\text{in}}-P_{\text{out}}).\\
    \end{split}
\end{equation}
We now subtract Eq.~\eqref{eq:p_jump_integral} from Eq.~\eqref{eq:mech_eq_integral} and obtain an integral expression for the pressure jump $\Delta P=P_{\text{in}}-P_{\text{out}}$ across the interface
\begin{equation}
    \begin{split}
        \Delta P & =\frac{n}{R^{n}}\int_{R_{\text{in}}}^{R_{\text{out}}}\text{d}r\;r^{n-1}[P_{\text{J}}(r;R)-P_{\text{N}}(r)]\\
        & +\frac{d-1}{R^{n}}\int_{R_{\text{in}}}^{R_{\text{out}}}\text{d}r\;r^{n-1}[P_{\text{N}}(r)-P_{\text{T}}(r)].\\
    \end{split}
\end{equation}
Finally, we set $n = d - 1$, thus eliminating the normal component of the pressure tensor $P_{\text{N}}$, and equate to the generalized Laplace law (cf. Eq.~\eqref{eq:LaplaceGen}) 
yielding
\begin{equation}
\begin{split}\Delta P & =\frac{d-1}{R^{d-1}}\int_{R_{\text{in}}}^{R_{\text{out}}}\text{d}r\;r^{d-2}[P_{\text{J}}(r;R)-P_{\text{T}}(r)]\\
 & =\frac{(d-1)\sigma[R]}{R}+\left[\frac{\text{d}\sigma}{\text{d}R}\right].
\end{split}
\end{equation}
It is possible to extract the expressions for $\sigma[R]$ and $[\mbox{d}\sigma/\mbox{d}R]$~\cite{RowlinsonWidom82} obtaining for $d = 3$
\begin{equation}\label{eq:gen_sigma}
    \sigma[R]=\int_{0}^{+\infty}\text{d}r\;\left(\frac{r}{R}\right)^{2}[P_{\text{J}}(r;R)-P_{\text{T}}(r)],
\end{equation}
\begin{equation}\label{eq:gen_d_sigma}
    \left[\frac{\text{d\ensuremath{\sigma}}}{\text{d}R}\right]=-\frac{2}{R^{3}}\int_{0}^{+\infty}\text{d}r\;r(r-R)[P_{\text{J}}\left(r;R\right)-P_{\text{T}}\left(r\right)],
\end{equation}
where we took the limits $R_{\text{in}} \to0$ and $R_{\text{out}}\to\infty$. We evaluate Eq.~\eqref{eq:gen_sigma} by means of the SC lattice pressure tensor in Eq.~\eqref{eq:sm:pt}, integrating along the $x$ axis so that $P_{\text{N}} = P^{xx}$ and $P_{\text{T}} = P^{yy} = P^{zz}$. Once $\sigma[R]$ is evaluated we obtain the value of $R_s$ by interpolating the position of the minimum. We wish to stress that in the derivation of Eq.~\eqref{eq:gen_sigma} the only hypothesis that has been used is that of mechanic equilibrium. It is possible to calculate an analytical expression for $\sigma[R]/\sigma_s$: recasting Eq.~\eqref{eq:LaplaceGen} as $R^{2}\Delta P=\text{d}\left[R^{2}\sigma\left[R\right]\right]/\text{d}R$, we integrate from $R_s$ to $R$ and obtain~\cite{RowlinsonWidom82} the expression $\frac{\sigma\left[R\right]}{\sigma_{s}}=\frac{1}{3}\left(\frac{R_{s}}{R}\right)^{2}+\frac{2}{3}\frac{R}{R_s}$, which in~\cite{Troster2011} is referred to as ``universal'', i.e. not depending on temperature or on the droplet/bubble size, mirroring that $\sigma[R]$ depends on the arbitrary value of $R$.
\begin{figure}[!t]
    \includegraphics[scale = 0.45]{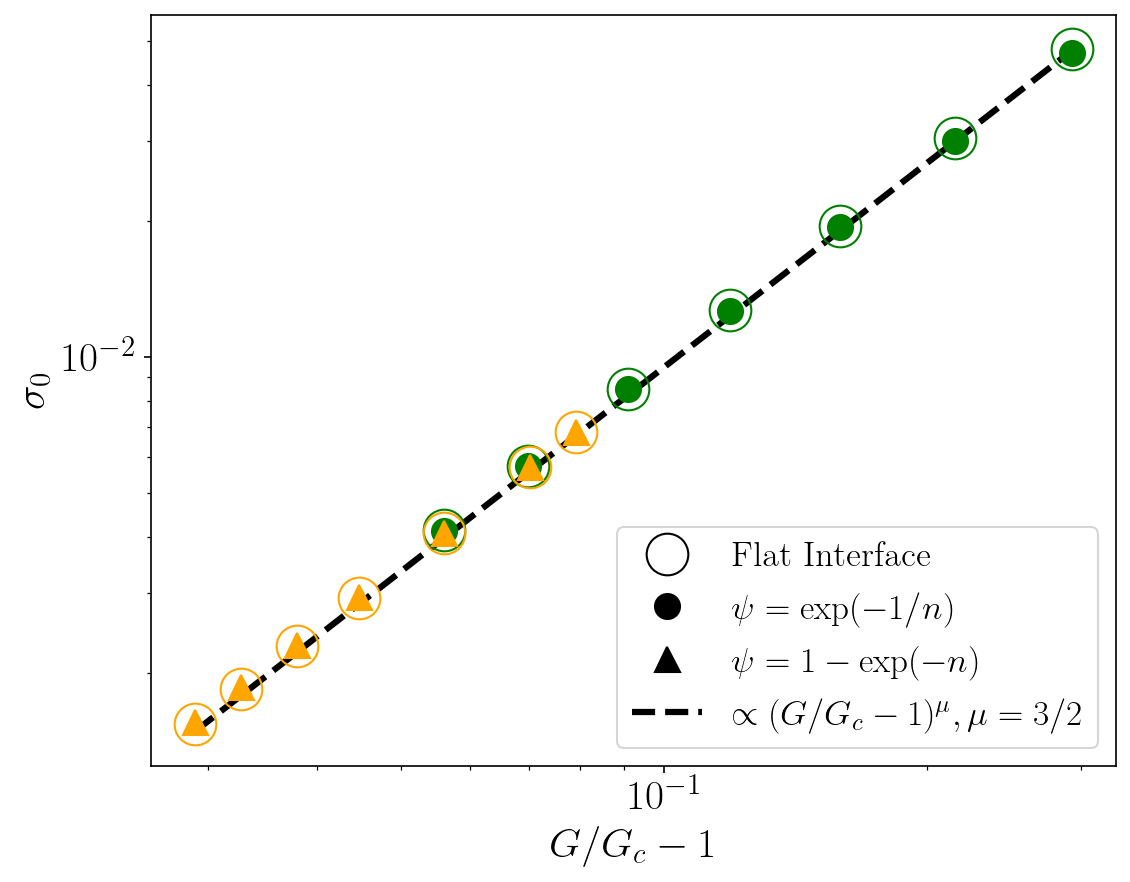}
    \caption{Value for the surface tension as a function of the dimensionless coupling $G/G_c - 1 \in \{0.029, 0.033, 0.038, 0.045,$ $ 0.056, 0.070, 0.079,$ $0.091, 0.119, 0.159, 0.215, 0.293\}$. Triangles and filled circles represent the data obtained from the interpolation of $R_s \cdot \Delta P / 2$ in the limit $R_s^{-1} \to 0$, for two different pseudo-potential functions $\psi$ (cf. Fig.~\ref{fig:3}$(a)$). Empty circles indicate the value numerically computed from the flat-interface simulations. The dashed line represent a power-law scaling with the mean-field exponent $\mu = 3/2$~\cite{Rowlinson1984}.}
    \label{fig:sigma_0}
\end{figure}
\begin{figure}
    \centering
    \includegraphics[scale = 0.45]{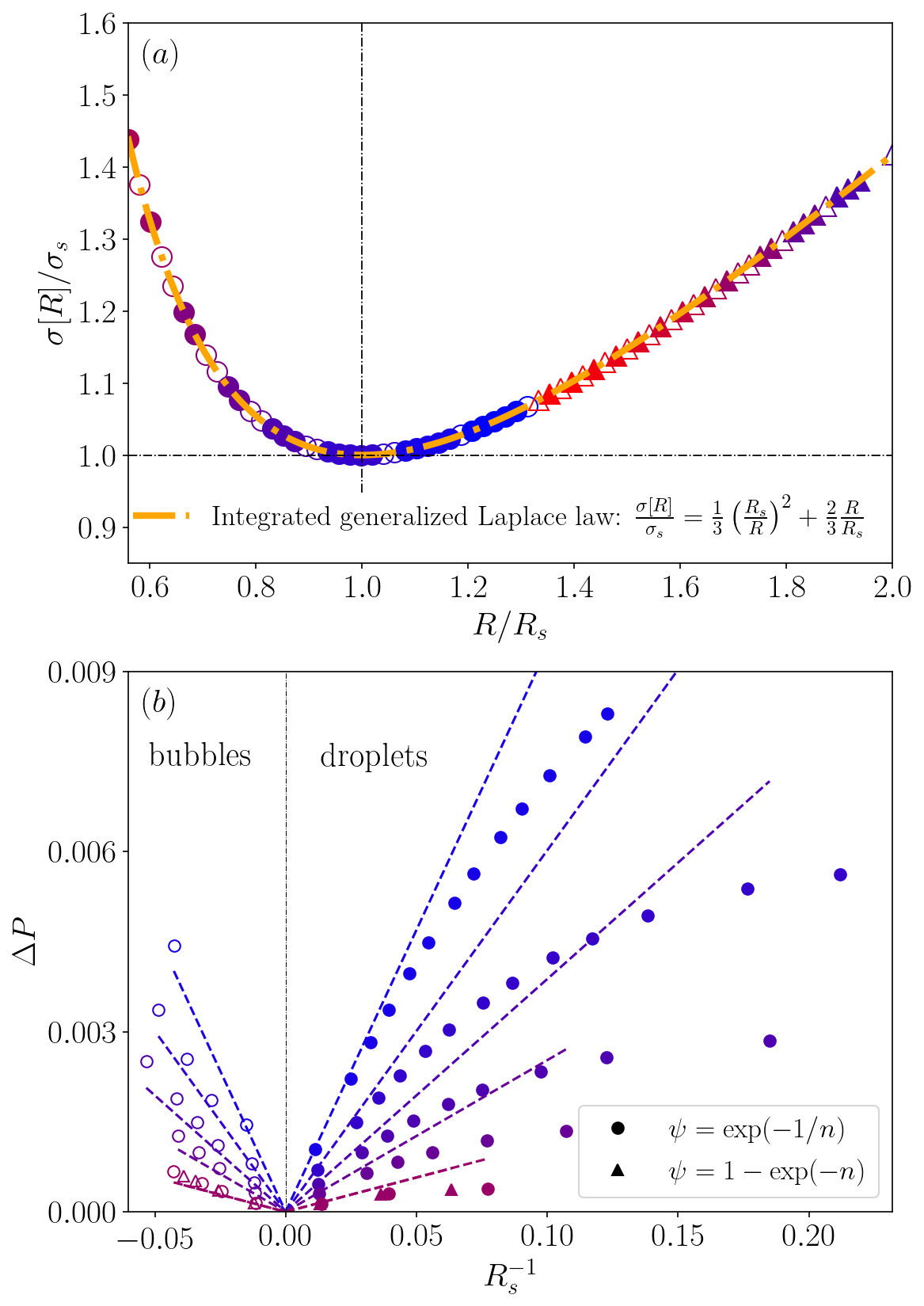}
    \caption{Panel $(a)$: surface tension at an arbitrary dividing surface $\sigma[R]$ normalized by the minimum $(R_s, \sigma_s)$ for all simulations: data collapse on a master curve (see text for details) independently on  the droplet/bubble size, coupling constant and equation of state. Panel $(b)$: dashed lines indicate the Laplace law using the values of $\sigma_0$ reported in Fig.~\ref{fig:sigma_0}, while the points represent the simulations data. Bubbles and droplets have opposite corrections with respect to the dashed lines. Colors correspond to the value of the dimensionless coupling $G/G_c - 1 \in \{0.070, 0.119, 0.159, 0.215, 0.293\}$ from dark red to blue.}
    \label{fig:2}
\end{figure}

\section{Results}\label{sec:res}
The simulations source code can be found on GitHub \href{https://github.com/lullimat/idea.deploy}{https://github.com/lullimat/idea.deploy}~\cite{sympy, scipy, numpy0, numpy1, scikit-learn, matplotlib, ipython, pycuda_opencl}. A Jupyter notebook~\cite{ipython} is available from the ``idea.deploy" framework to reproduce the results reported in this paper. We simulate three-dimensional droplets and bubbles in a cubic system of linear size $L$ with periodic boundary conditions using the D3Q19 discrete velocity set with $c_s^2=1/3$~\cite{kruger2017lattice, succi2018lattice} (see Fig.\ref{fig:sketch}). We adopt two possible definitions for the pseudo-potential function, namely $\psi = \exp(-1/n)$ and $\psi = 1 - \exp(-n)$~\cite{ShanChen94, Sbragaglia07}. The range of dimensionless coupling constants $G/G_c - 1 \in \{0.029, 0.033, 0.038, 0.045, 0.056, 0.070, 0.079,$ $0.091, 0.119, 0.159, 0.215, 0.293\}$ where $G_c c_s^2= -2.463$ and $G_c c_s^2 = -1.333$ for $\psi = \exp(-1/n)$ and $\psi = 1 - \exp(-n)$, respectively. The value of $L$ is chosen to be an odd number so that the center of mass of the system exactly falls on the coordinates of a node. The simulated system sizes are $L\in \{41,  43,  47, 51,  55,  61,  67, 77, 87, 103, 123, 157, 213, 335\}$. The radial density field $n(r)$ is initialized to the following profile
\begin{equation}
  n(r, R) = \frac{1}{2}(n_{\text{in}} + n_{\text{out}}) - \frac{1}{2}(n_{\text{in}} - n_{\text{out}})\tanh(r - R),
\end{equation}
where the inner $n_{\text{in}}$ and outer $n_{\text{out}}$ densities are initialized to the equilibrium values of the gas $n_g$ and liquid $n_l$ for a flat interface~\cite{Shan08}, for bubbles and droplets accordingly. The initial value of the radius is set to maintain a fixed aspect ratio for all simulations as $R = L/4$. The radial coordinate $r$ is computed taking the center of the system as the origin. The values $P_{\text{in}}$ and $P_{\text{out}}$ are computed in the middle of the system $(\lfloor L/2 \rfloor, \lfloor L/2 \rfloor, \lfloor L/2 \rfloor)$ and at the farthest corner $(L - 1, L - 1, L - 1)$, respectively. The outcome of the simulations is analyzed only if all the coordinates of the center of mass lie within a distance of $10^{-3}$ from the center of the domain. We use two convergence criteria for the simulations, both comparing quantities at a time distance $\delta t = 2^{11}$: i) we consider the relative variation of the $\Delta P$ with respect to the previous configuration, and when the latter is such that $|\Delta P(t) - \Delta P(t + \delta t)|/\Delta P(t) < 10^{-5}$ the simulation is considered as converged; ii) we consider the magnitude $\delta u$ of the spatial average of the difference between the components of two velocity fields, $\delta u=L^{-3}\sum_{\textbf{x}}\sum_{\alpha}|u^{\alpha}(\textbf{x},t+\delta t)-u^{\alpha}(\textbf{x},t)|$ so that the simulation is considered as converged when $\delta u< 10^{-12}$. Meeting only one of the two criteria is enough to finalize the simulation.
\begin{figure}[!t]
    \includegraphics[scale = 0.45]{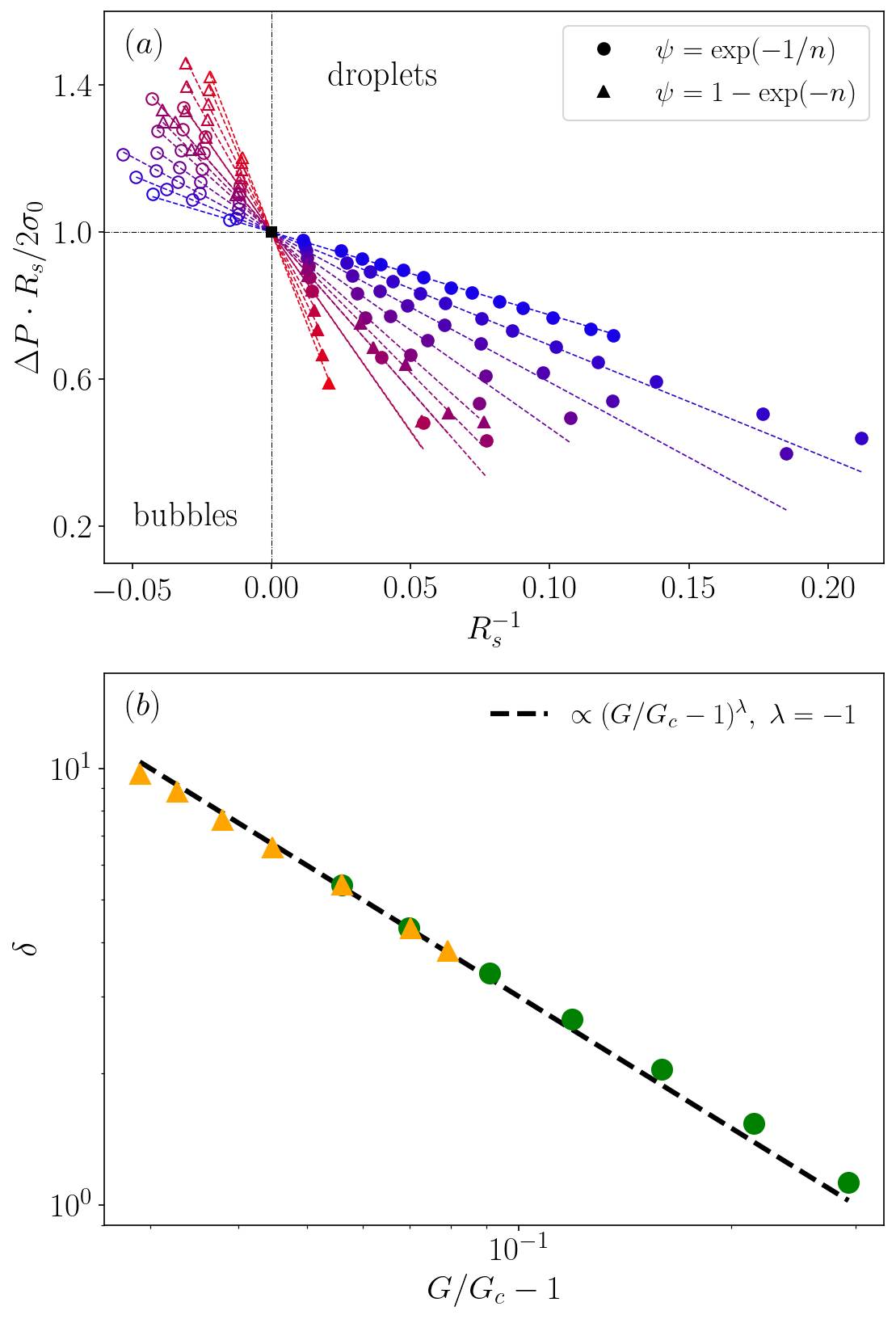}
    \caption{Panel $(a)$: Surface tension at the surface of tension computed from the simulations by means of the Laplace law $\sigma_s(R_s) = \Delta P \cdot R_s / 2$. Colors correspond to different values of the dimensionless coupling $G/G_c - 1 \in \{0.029, 0.033, 0.038, 0.045, 0.056$ $, 0.070, 0.079, 0.091, 0.119, 0.159, 0.215, 0.293\}$ from red to blue. Panel $(b)$: results for the Tolman length $\delta$ estimated by the fits in panel $(a)$ as a function of the dimensionless coupling $G/G_c - 1$: data for different choices of $\psi$ fall onto a universal curve which is well approximated by a power law with exponent $\lambda \simeq -1$.}
    \label{fig:3}
\end{figure}
The set of simulations for the flat interface has been performed on a two-dimensional domain using the D2Q9 discrete velocity set with $c_s^2=1/3$~\cite{kruger2017lattice, succi2018lattice} and the forcing weights are obtained by the projection of the three-dimensional case, i.e. $W(1) = 1/3$ and $W(2) = 1/12$. The domain sizes are $L_x = 100, L_y = 4$ for all simulations and the density profile is initialized according to
\begin{equation}
  \begin{split}n\left(x,x_{0},w\right)= & \frac{1}{2}\left(n_{l}+n_{g}\right)\\
- & \frac{1}{2}\left(n_{l}-n_{g}\right)\tanh\left[x-\left(x_{0}-\frac{w}{2}\right)\right]\\
+ & \frac{1}{2}\left(n_{l}-n_{g}\right)\left\{ \tanh\left[x-\left(x_{0}+\frac{w}{2}\right)\right]+1\right\},
\end{split}
\end{equation}
where $x_0$ is the center of the strip and $w = L_x/2$ its width.

We report in Fig.~\ref{fig:sigma_0} the value for the surface tension $\sigma_0$ in the flat interface limit. Full symbols represent the interpolation of the droplets/bubbles data for $\sigma_s(R_s) = R_s \cdot \Delta P / 2$ in the limit $R_s^{-1} \to 0$. Such values of $\sigma_0$ are used in Fig.~\ref{fig:3} as normalization constant. Empty circles represent the results obtained from the flat-interface simulations by numerically computing the integral
\begin{equation}
    \sigma_0 = \int_{L_x/2}^{L_x} \mbox{d}x [P_{\text{N}}(x) - P_{\text{T}}(x)]
\end{equation}
where $P_{\text{N}}(x) = P^{xx}(x)$ and $P_{\text{T}}(x) = P^{yy}(x)$ have been obtained from the lattice pressure tensor~\eqref{eq:sm:pt} using the two-dimensional values of the weights $W(1) = 1/3$ and $W(2) = 1/12$. The scaling with respect to the dimensionless coupling $G/G_c - 1$ matches the mean-field case with exponent $\mu = 3/2$~\cite{Rowlinson1984, MAYER_2004}. Different choices for the pseudo-potential function, yielding different equations of state, result in the same scaling law and the same prefactor thus implying that the results belong to the mean-field \emph{universality} class. In particular, this result allows to describe both set of data in terms of a single reduced-coupling scale $G/G_c - 1$.

Figure~\ref{fig:2}$(a)$ displays the values for $\sigma[R]$ (see Eq.~\eqref{eq:gen_sigma}) obtained from the simulation data, superposing to the expected integrated result for $\sigma[R]/\sigma_s$, for bubbles and droplets of different sizes, with different equations of state $P(n)$~\eqref{eq:bulkp} and at different temperatures. Hence, based on the derivation in~\cite{RowlinsonWidom82} and using the SC lattice pressure tensor~\cite{Shan08} we obtain a result that is compliant with the thermodynamics of curved interfaces~\cite{GibbsCollected1948, RowlinsonWidom82} allowing us to estimate $R_s$. Notice that most of MD works rely on the use of the equimolar radius $R_e$, implying in three dimensions the cancellation of the second order curvature corrections in $\sigma(R)$~\cite{Blokhuis1992Rigidity}, with the exception of~\cite{Rao1979} that applies the so-called mechanical definition of $\delta$ for a flat interface. In order to find the value of $R_s$ the authors of~\cite{Troster2011} estimate the minimum of $\sigma[R]$ from the statistical average of an excess free energy normalized to the area of spherical volumes of varying size by means of MC simulations, with the need to keep in check finite-size corrections to the statistics. Both MD and MC have in common the necessity of averaging quantities over thermal fluctuations, which however is not required in the present LBM simulations.
\begin{figure}[!t]
    \includegraphics[scale = 0.45]{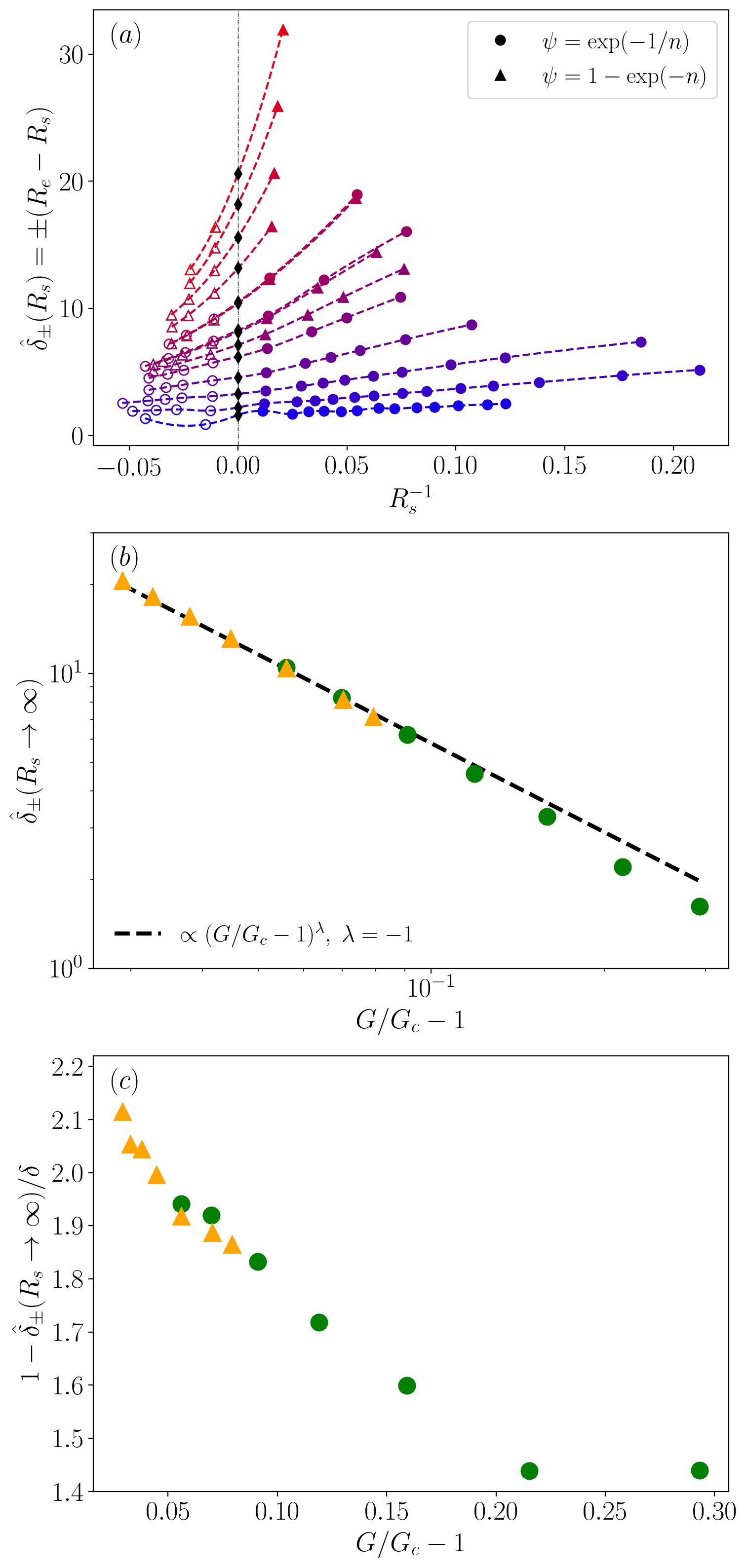}
    \caption{Panel $(a)$: Curvature dependence of the alternative definition of the Tolman length $\hat{\delta}_\pm(R_s) = \pm (R_e - R_s)$, reporting in black diamonds the extrapolated value in flat interface limit $R_s^{-1}\to 0$ for different values of the dimensionless coupling $G/G_c - 1 \in \{0.029, 0.033, 0.038, 0.045,$ $ 0.056, 0.070, 0.079,$ $0.091, 0.119, 0.159, 0.215, 0.293\}$ with colors ranging from red to blue. Panel $(b)$: Scaling of the flat interface limit value against $G/G_c - 1$. Panel $(c)$: Quantitative comparison against the definition from Eq.~\eqref{eq:Laplace2nd}.}
    \label{fig:sm:2}
\end{figure}

In Fig.~\ref{fig:2}$(b)$ we report the data points $(R_s^{-1}, \Delta P)$ and compare to the Laplace law, considering bubbles as having negative curvature. The slope of the dashed lines is given by $2\sigma_0$ and deviations from the flat interface limit appear with opposite sign for bubbles and droplets. Next, we analyze the corrections: in Fig.~\ref{fig:3}$(a)$ we show the data for the surface tension at the surface of tension estimated from $\Delta P$ and $R_s$, i.e. $\sigma_s(R_s) = \Delta P \cdot R_s/2$~\footnote{The values of $\sigma_s = \sigma[R_s]$ estimated from the minimum of $\sigma[R]$ match those obtained from $\Delta P$ with a relative error of order $10^{-3}$.}, normalized by the flat interface value $\sigma_0$, i.e. we analyze the $yz$ projection of Fig.~\ref{fig:sketch}$(a)$ for different temperatures. We first determine the value of $\sigma_0$ by interpolating the data for $\sigma_s(R_s)$ as a function of $R_s^{-1}$ in the limit $R_s^{-1}\to 0$. Such values match those computed from flat interface simulations (see Fig.~\ref{fig:sigma_0}). Then we fit the corrections which are well approximated by linear functions (see Eq.~\eqref{eq:Laplace2nd}) reported in dashed. We estimate the Tolman length from the lines slope which is equal to $-2\delta$. In Fig.~\ref{fig:3}$(b)$ we report the values of $\delta$ as a function of the dimensionless coupling $G/G_c - 1$: data for different $\psi$ lie on the same curve which is well approximated by a power law with exponent $\lambda = -1$. It is possible to compute this value of the exponent from the expression $\lambda = -\nu -\beta$~\cite{Blokhuis1992} when inserting the mean-field values of the exponents $\beta = 1/2, \nu = 1/2$, which characterize the critical behavior of the order parameter (liquid-gas density difference) and of the correlation length, respectively~\cite{MAYER_2004}. The latter expression for $\lambda$ has been derived by Blokhuis and Bedeaux in~\cite{Blokhuis1992} from the expansion in $R^{-1}$ of $\Delta P$ for a spherical surface, which they could match with the expansions for $\sigma[R]$ and $\sigma'[R]$ obtained from thermodynamic arguments in~\cite{Blokhuis1992Rigidity}. In particular they found that the flat-interface definition~\cite{Tolman1949} $\delta = z_e - z_s$, where $z_e$ and $z_s$ are the positions of the equimolar surface and of the surface of tension respectively, needs to be modified into $\delta' = \delta + A$, obtained as the infinite radius limit of the curvature expansion. While it is known that $\delta$ has a zero mean-field exponent it is $A$~\cite{Blokhuis1992} that yields the singular behavior. Such a modified expression for the Tolman length of a flat surface has also been used in MD studies~\cite{Lei2005}. Theoretical mean-field works report a negative sign for $\delta$ leveraging, however, the flat-interface definition $\delta = z_e - z_s$ which can be modified as discussed above~\cite{Blokhuis1992} thus possibly changing the sign~\cite{RowlinsonWidom82}. Lattice-gas results~\cite{Troster2011} as well as a recent molecular simulation~\cite{Menzl2016}, report a positive sign as we find in the present work. Indeed, any further quantitative comparison between LBM and MD would require a direct mapping between the pseudo-potential function $\psi$, defining the lattice SC force, and the pair interaction potential used in MD, which however is a rather delicate task. Indeed, in the recent years, there have been some progress in mapping MD onto equivalent lattice Boltzmann schemes~\cite{Parsa_2017} although being limited to the single-phase case.

We report now a detailed comparison of the definition of the Tolman length adopted so far, against another possible choice $\hat{\delta}_\pm(R_s) = \pm (R_e - R_s)$, with the plus and minus signs for droplets and bubbles respectively, adopted in~\cite{Troster2011}. The equimolar radius $R_e$ is defined by solving the equation $M = 4\pi R_{e}^{3}\,n_{\text{in}}/3+(V-4\pi R_{e}^{3}/3)\,n_{\text{out}}$ where $M$ is the total mass of the system of volume $V = L^3$. This equation is equivalent to the request of vanishing adsorbance~\cite{RowlinsonWidom82} as a function of the arbitrary dividing surface. In Fig.~\ref{fig:sm:2}$(a)$ we report the size dependence of $\hat{\delta}_\pm(R_s)$ for different values of the coupling $G$ reporting in black diamonds the extrapolated value in the flat interface limit $R_s^{-1}$. In Fig.~\ref{fig:sm:2}$(b)$ the latter are plot against the dimensionless coupling $G/G_c - 1$ on a log-log scale showing the same scaling relation obtained for the alternative definition $\delta$. Finally in Fig.~\ref{fig:sm:2}$(c)$ we report a quantitative comparison between the two different definitions.

\section{Conclusions}\label{sec:conslusions}
In conclusion, we demonstrate the ability of the multi-phase Shan-Chen LBM to capture relevant features of the curvature corrections to the surface tension: we find a temperature-dependent Tolman length $\delta$ displaying a power-law behavior near the critical point. Furthermore, $\delta$ shows a universal scaling for different equations of state. The advantage of this approach is manifold: i) the thermodynamic properties of the interfaces are emergent, as in MD, but from an underlying simplified lattice dynamics, ii) the intrinsic hydrodynamic compliance of LBM is unprecedented in the previous simulation literature, opening a new direction in which to study systematically the role of curvature corrections in more complex hydrodynamic regimes and iii) the contained computational cost allows to explore a broad parameter space. Future work will probe the possibility of using more refined versions of the Shan-Chen model and tune the different curvature coefficients similarly to what has been previously done with the surface tension~\cite{Sbragaglia07} and disjoining-pressure~\cite{Benzi_2009_1}, as well as refining mathematical control of the model. Finally, we shall consider hydrodynamic fluctuations compliant with thermodynamics by extending to the SC multi-phase model the works~\cite{Gross_2010, Belardinelli15} which leverage the Multi-Relaxation time collisional operator. By doing so, we would be able to study the effects of the Tolman length on the homogeneous nucleation rates along the lines of~\cite{Menzl2016, Aasen2020}. The simulations source code and a Jupyter notebook to reproduce all the results and figures can be found on GitHub \href{https://github.com/lullimat/idea.deploy}{https://github.com/lullimat/idea.deploy}.

\begin{acknowledgements}
The authors wish to thank {\O}ivind Wilhelmsen for useful discussion. Luca Biferale thankfully acknowledges the hospitality from the Department of Mechanics and Aerospace Engineering of Southern University of Science and Technology. This work was supported by National Science Foundation of China
 Grants~12050410244, 91741101 and 91752204, by Department of Science and Technology of
 Guangdong Province Grant No.~2019B21203001, Science and Technology
 Innovation Committee of Shenzhen Grant No.~K19325001, and from the European Research Council (ERC) under the European Union’s Horizon 2020 research and innovation programme (grant agreement No 882340).
\end{acknowledgements}

\bibliography{Biblio}
\end{document}